\documentclass[%
 reprint,
superscriptaddress,
nofootinbib,
 amsmath,amssymb,
 aps,prd,
]{revtex4-2}
\usepackage{amsmath}
\usepackage[section]{placeins}
\usepackage[utf8]{inputenc}
\usepackage[english]{babel}
\usepackage{float}
\usepackage{isotope}

\usepackage[dvipsnames]{xcolor}

\usepackage{soul}
\usepackage{graphicx}
\usepackage{dcolumn}
\usepackage{bm}

\usepackage{graphicx}
\usepackage{dcolumn}
\usepackage{bm}

\usepackage[draft, inline, nomargin]{fixme}
\fxsetup{theme=color, mode=multiuser}

\FXRegisterAuthor{bc}{1}{\color{red}BC}
\FXRegisterAuthor{ag}{2}{\color{blue}AG}
\FXRegisterAuthor{ph}{3}{\color{magenta}PH}

\newcommand{\y}{$^{90}$Y}
\newcommand{\srnine}{$^{90}$Sr}
\newcommand{\ce}{$^{144}$Ce}
\newcommand{\pr}{$^{144}$Pr}
\newcommand{\ru}{$^{106}$Ru}
\newcommand{\rh}{$^{106}$Rh}

\begin{document}


\title{Cerium ruthenium low-energy antineutrino measurements for safeguarding military naval reactors}

\author{Bernadette K. Cogswell}
\email{bkcogswell@vt.edu}

\affiliation{Center for Neutrino Physics, Physics Department, Virginia Tech, Blacksburg, VA 24061}

\author{Patrick Huber}
\email{pahuber@vt.edu}
\affiliation{Center for Neutrino Physics, Physics Department, Virginia Tech, Blacksburg, VA 24061}

\date{\today}%

\begin{abstract}
The recent agreement to transfer nuclear submarine reactors and technology from two nuclear-weapon states to a non-nuclear-weapon state (AUKUS deal) highlights an unsolved problem in international safeguards: how to safeguard naval reactor fuel while it is on-board an operational nuclear submarine. Proposals to extend existing safeguards technologies and practices are complicated by the need for civilian international inspectors to gain access to the interior of the submarine and the reactor compartment, which raises national security concerns. In this paper we show that implementing safeguards on submarine propulsion reactors using a
low-energy antineutrino reactor-off  method, between submarine patrols, can by-pass the need for on-board access all together. We find that, using inverse beta decay (IBD), detectors can achieve a timely
and high level of assurance that a submarine's nuclear core has not been diverted (mass of around 100\,kg) nor its enrichment level changed (mass of around 10\,tons).

\end{abstract}

\maketitle

\section{\label{sec:introduction}Introduction}

Highly-enriched uranium (HEU) for military naval reactors poses
specific and significant challenges for
non-proliferation~\cite{shea,nti}. The Treaty on the Non-proliferation
of Nuclear Weapons (NPT) allows for the withdrawal of HEU from the
civilian realm, and thus from safeguards, to move it to military
non-explosive uses, such as naval reactors. This is a particular concern because
it could create a pathway to nuclear weapons for non-weapon state
parties without the risk of detection via international safeguards.
To date, only states with nuclear weapons have deployed nuclear
powered submarines, which rendered the related proliferation concerns
theoretical. Brazil's plans to build a nuclear powered submarine make
this a more concrete concern, but progress towards actual deployment
has been slow~\cite{brazil}. With the recently announced
Australia-U.K.-U.S. (AUKUS) agreement~\cite{AUKUS} to transfer nuclear
submarines from two NPT weapon states (the U.S. and U.K.) to an NPT
non-weapon state (Australia) the question of how to implement naval reactor safeguards, in order to detect and deter the diversion of nuclear weapons-usable material,
has become urgent.

One approach is to phase-out reliance on HEU in naval propulsion
programs~\cite{shea,nti,MaHippel}. Both the U.S. and U.K. navies
exclusively employ HEU above 90\% enrichment\footnote{Generally,
uranium of enrichment less than 20\% is referred to as LEU and
everything above 20\% enrichment as HEU.} in submarine propulsion and,
in particular, the U.S. Navy has not yet pursued switching to low-enriched
uranium (LEU)~\cite{navy1995,navy2014}. However, regardless of if HEU
or LEU is in use in naval reactors, safeguards need to be
considered. Usually the need for keeping military secrets is cited to
counter this possibility, however, some
studies~\cite{shea,Philippe:2014,costa} make the case that the need
for secrecy does not preclude meaningful safeguards, if managed access
to the reactor is available. Here, we will consider the case of no
access to the reactor (or even access on-board the submarine), which
goes beyond what has been considered in the literature so far. We
propose to use offline, {\it i.e.}, during reactor shutdown,
neutrino\footnote{In this letter we are only concerned with
antineutrinos and, hence, will use the term neutrino for these as
well.} measurements in port to ascertain the presence of a nuclear reactor. Neutrino emissions after reactor shutdown have been
considered in the context of spent nuclear fuel, see {\it e.g.},
Ref.~\cite{Brdar2017}, and have been previously observed by the Double
Chooz experiment~\cite{DoubleChooz:2020vtr}. The new technique we
propose here is based on the observation of CErium RUthenium Low
Energy ANtineutrino (CeRuLEAN) emissions.

For naval reactors the main safeguards objective is the
verification that the vessel is nuclear-powered, {\it i.e.}, that the
reactor is still present and has not been replaced with a non-nuclear
energy source. Even for a would-be proliferator without strong domestic
nuclear reactor expertise this presents a possible diversion
pathway. In the literature a technical proposal for implementation of
this objective has emerged: so-called flux tabs~\cite{shea}. Flux tabs
are made of a material which gets activated, {\it i.e.}, becomes
radioactive, under neutron irradiation. The idea is to place these flux
tabs close to (but outside of) the reactor, in areas that receive a
significant flux of fast neutrons from the fissions going on in the
reactor. The level of radioactivity in these flux tabs is proportional to the energy the reactor has produced. The problem is that
these flux tabs need to be installed close to the reactor and, hence, an
inspector also needs to obtain access to the vessel and get close to the
reactor. In order to protect classified information this would require
so-called "managed access", which is logistically complex to establish.

In this paper we will demonstrate that a small neutrino detector using
the CeRuLEAN technique can, however, serve as a direct equivalent of these flux
tabs \emph{without} the need for any on-board access.

\section{Submarine and reactor considerations}

Power consumption for a submarine is dominated by the power used for
propulsion. Propulsion power is proportional to the drag; and that is
proportional to the third power of the speed, $v^3$. Assuming full
design reactor power $P_d$ at top speed we obtain the fractional power
usage as a function of speed,
\begin{equation}
    P(v)=P_d\left(\frac{v}{v_\textrm{max}}\right)^3\,,
\end{equation}
and we take $v_\textrm{max}=35\,$knots and
$P_d=150\,$MW$_\mathrm{th}$. This reactor power corresponds, for
example, to the S6G reactor employed in U.S. {\it Los Angeles} class
submarines, and is similar to the output of the S9G reactor used for
{\it Virginia} class vessels ~\cite{ragheb}, which make up roughly
three-quarters of the United States' submarine
force~\cite{ntiSubColl}. Furthermore, following Ref.~\cite{Ippolito}
we assume that the vessel spends two-thirds of the year at sea,
conducting two patrols of 4 months each per year, and spending one-third of the
year in port, representing a maximum possible in-port detector dwell
time of 4 months per calendar year. Taking an average cruise speed of 22\,knots, we find
that power consumption while at sea is 1/4\,P$_d$. There is also a small
level of energy consumption even at $v=0$, the so-called "hotel
load"~\cite{Ippolito} to sustain basic life-supporting crew
operations, which we will neglect. From these considerations it follows immediately that any attempt at
monitoring a naval reactor while at full or fractional cruising power
would require "drive-by" style monitoring, resulting in a very small
product of dwell time and reactor power. This makes online monitoring
(when the reactor is operating) of the reactor core via the usual neutrino-based
techniques~\cite{Detwiler:2002ym,Christensen:2014pva,Bernstein_2020}
impractical. Limited dwell time while the reactor is on also precludes
the use of more exotic techniques like the observation of breeding
neutrinos~\cite{Cogswell:2016aog} or non-linear effects related to
power density~\cite{Huber:2015ouo}.

Details of naval reactor design and operation are still largely
shrouded in secrecy, but many of the gross characteristics have become
available in the open literature. In particular, we will follow
Ref.~\cite{Ippolito}, which to our knowledge is the only detailed
reactor engineering study of submarine reactors which is openly
accessible. The goal of the study in Ref.~\cite{Ippolito} was to understand how
different fuel enrichment levels affect submarine reactor size and
lifetime. Five different reactor cores were studied and for the analysis
presented here we find that our results do not change appreciably between one of these five options. For the results presented in the main text we assume HEU cores, see appendix for details.

\section{Offline reactor monitoring}

A nuclear reactor emits neutrinos even after shutdown, stemming from
fission fragments with longer half-lives. Since we are interested in
using inverse beta decay as the detection reaction we look at isotopes with beta endpoint
energies in excess of 1.8\,MeV. It turns out that there are only four
decay chains that fulfill this criterion~\cite{Brdar2017} and have
lifetimes exceeding minutes; they are listed in
Tab.~\ref{tab:isotopes}.

\begin{table}[h]
    \centering
    \begin{tabular}{c|cccc}
    Parent & {\srnine} &{\ce}&{\ru}&$^{88}$Kr\\
    \hline
    
  Lifetime $\tau$ [d]&  15218 & 411 & 536 & 0.2\\
    Daughter & {\y}&{\pr}&{\rh}&$^{88}$Rb\\
    $Q_\beta$ [MeV] &2.28& 3.00& 3.54 &5.31\\
    $\sigma_\mathrm{IBD}\,[10^{-43}\,\mathrm{cm}^2]$&0.08& 0.45& 0.75&2.84\\
    \hline
  CFY $^{235}$U      & 0.057 & 0.055 & 0.004 & 0.035 \\
 CFY $^{239}$Pu &0.02 & 0.037 & 0.042 & 0.012 \\
         
    \end{tabular}
    \caption{Isotopes suitable for offline monitoring. Beta decay information from ENSDF~\cite{ensdf} and cumulative fission yield (CFY) information from JENDL-4.0~\cite{jendl}.}
    \label{tab:isotopes}
\end{table}

The decay energy $Q_\beta$ of the {\srnine}/{\y} chain is low and,
thus, its inverse beta decay cross section $\sigma_\mathrm{IBD}$ is
also low. The resulting suppression in the signal makes this chain not
ideal for measurement. The $^{88}$Kr/$^{88}$Rb chain has a lifetime
too short to effectively contribute to the detected rate. This leaves
{\ce}/{\pr} and {\ru}/{\rh} as the best candidates. The cumulative
fission yields (CFY) for $^{235}$U and $^{239}$Pu fission are quite
similar for {\ce}, but distinct by an order of magnitude for
{\ru}. Hence, measuring the ratio of {\ce} to {\ru} is a direct
measurement of the plutonium fission fraction, which is explored
in detail in the appendices. Note, the detection of neutrinos from a purpose-made {\ce} source
has been proposed for sterile neutrino searches with the Borexino
detector~\cite{Borexino:2013xxa}.

The lifetimes for {\ce} and {\ru} are comparable to the patrol and
off-duty periods postulated for a submarine and, thus, we need to
perform a calculation of their abundance $n_i$ with $i$ being
one of the four isotopes:
\begin{eqnarray}
    n'_i(t)&=&-\frac{1}{\tau_i} n_i(t)\\ &+& CFY_i^\mathrm{U235}f_\mathrm{U235} F(t) +  CFY_i^\mathrm{Pu239}f_\mathrm{Pu239} F(t)\,,\nonumber
\end{eqnarray}
where $f$ represents the fission fraction for $^{235}$U or $^{239}$Pu,
as applicable, $f_\mathrm{U235}=1-f_\mathrm{Pu239}$, and $F(t)$ is the
overall fission rate corresponding to 31.5\,MW divided by the energy
per fission (200\,MeV) for the four months the vessel is at sea with
$F(t)=0$ for the two months in port.

For the detector monitoring set up, we assume a 1\,ton detector at
5\,m from the center of the reactor and a dwell time of 2 months. In
terms of total event numbers, we expect for a fresh core between
108 and 164 events for an $f_\mathrm{Pu239}$ of 0 or 1, respectively. After about
5 calendar years of operation these numbers reach their asymptotic
value of 308--512 events.

\begin{figure*}[t!]
    \centering
    \includegraphics[width=0.66\textwidth]{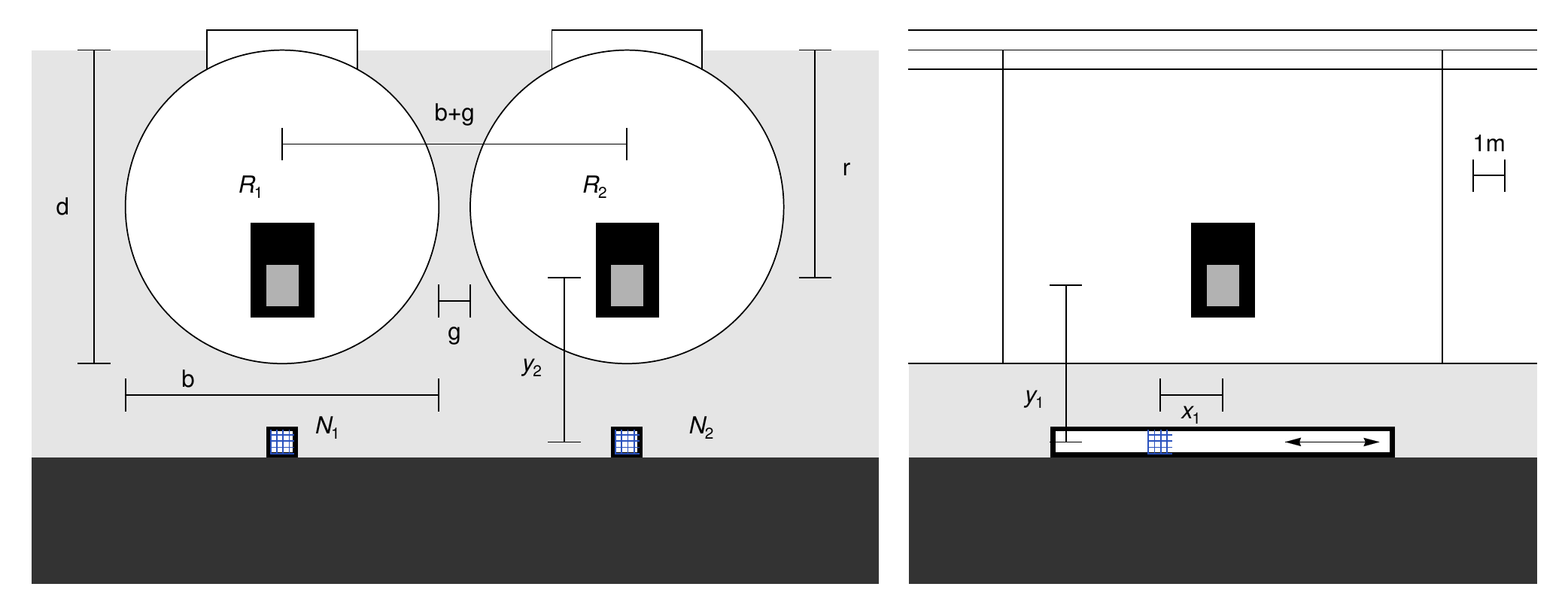}%
   
    \caption{Cross section (left panel) and side view (middle panel) of two submarines at their berthing site. $R_1$ denotes reactor of vessel 1 and $N_1$ denotes the neutrino detector for that vessel. $d$ is the draft, $b$ is the beam of the vessel and $r$ denotes how far the center of the reactor core is below the water line. $g$ describes the gap between vessels while moored. $y_1$ is the vertical distance between the center of the reactor core $R_1$ and the neutrino detector $N_1$. $x_1$ is the horizontal, adjustable distance between $R_1$ and $N_1$. Water is depicted as light gray and the harbor floor as dark grey. Drawing is to scale, reactor size and position are notional.}
    \label{fig:port}
\end{figure*}

A detector placement below the submarine, as modeled here and shown in
Fig.~\ref{fig:port}, corresponds to about 15--20 meters water
equivalent overburden, given a distance from the reactor core of 5--10\,m
and that the typical attack submarine has a draft (vertical distance
from the waterline to the bottom of the hull) of about 10\,m, with the
reactor being somewhat below the center line of the
submarine~\cite{friedman,polmar}. This is similar to the 20 meter
water equivalent overburden of the NEOS neutrino
experiment~\cite{NEOS:2016wee}. NEOS is a 1\,ton, single volume
detector and observes around 80 background events per day, about half
of these in the relevant region below 4\,MeV. This corresponds to
about 60 events per 0.1\,MeV in 60 days for a 1\,ton detector. The
background stems from three sources: accidental coincidences, which
scale with detector volume; fast neutrons, which also scale with
detector volume; and cosmogenic beta-delayed neutron emitters
($^9$Li), which scale with the muon rate and, hence, the detector
surface area.

In a segmented detector~\cite{Ashenfelter:2018iov,Haghighat:2018mve}
both volume and surface-scaling backgrounds are sharply reduced: a
true inverse-beta decay event occupies a certain volume enveloped by a
certain surface. For example, in the CHANDLER~\cite{Haghighat:2018mve}
design the whole detector consists of cubes measuring 6.2\,cm in each
direction. The coincidence volume for an inverse beta decay event is
$3\times3\times 3=27$ cubes and the corresponding top surface is
$3\times 3=9$ cube faces. Hence, a 1\,ton detector would consist of
roughly 4\,100 cubes and have a top surface area of about
$4\,100^{2/3}\simeq 260$ cube faces. Therefore, the rate of
volume-scaling backgrounds, like accidentals, is reduced by a factor
$4\,100/27\simeq 150$ and for surface-scaling backgrounds the
reduction factor is $260/9\simeq 30 $. Thus, for the 1\,ton detector
considered here, this yields scaled background rates in the range of
0.04 to 2 per 60-day period and per 0.1\,MeV.  The
accidental background scaling has been experimentally
verified~\cite{Haghighat:2018mve}. However, whether or not the muon
related backgrounds indeed scale with segmentation as described has
yet to be tested, but remains a reasonable assumption. For the remainder of this study we will use an intermediate value 0.5 backgrounds events per 60 day period and 0.1\,MeV, corresponding to 25 events in 60 days.

A neutrino measurement allows for an effective information barrier:
presumably, the owner of the vessel would want to keep the reactor full power equivalent days (FPED) during
a patrol secret, since this number would allow inferences to
be made about the patrol distance. The strength of the {\ce} signal, being only a weak function of
$f_\mathrm{Pu239}$, would allow a fairly accurate measurement of the
accumulated FPED of the vessel in the few hundred days prior to the measurement being taken. However, by
varying the reactor-detector distance\footnote{Using ``false''
distances was used as a blinding scheme in the Day Baya experiment to
prevent analysis bias~\cite{DayaBay:2012fng}.} (via changing $x_1$ within a fixed and known range, see Fig.~\ref{fig:port}) the owner of the vessel
can effectively erase this signature: for a high FPED patrol the
detector would move further away and for a low FPED patrol the
detector moves closer. The actual reactor-detector distance can easily
be concealed from the inspecting party, by design, see Fig.~\ref{fig:port}.
We call the corresponding parameter the {\it power masking factor} $\xi=(d_\mathrm{max}/d_0)^2$, defining the degree to which FPED information can be hidden. The actual value is determined by the extent to which the owner of the vessel wants to keep the true reactor usage on a patrol secret. Once this total
signal strength information has been erased, the only remaining
information is that (1) there is a significant neutrino flux, which
encodes the fact that a reactor with a certain minimum FPED is
present, and (2) the ratio of {\ce} to {\ru} neutrinos, which encodes
the accumulated fission fraction $f_\mathrm{Pu239}$. The latter
information would require a detector with a mass of around 10\,tons to be effectively
usable, see appendices.

\begin{figure}
    \centering
    \includegraphics[width=\columnwidth]{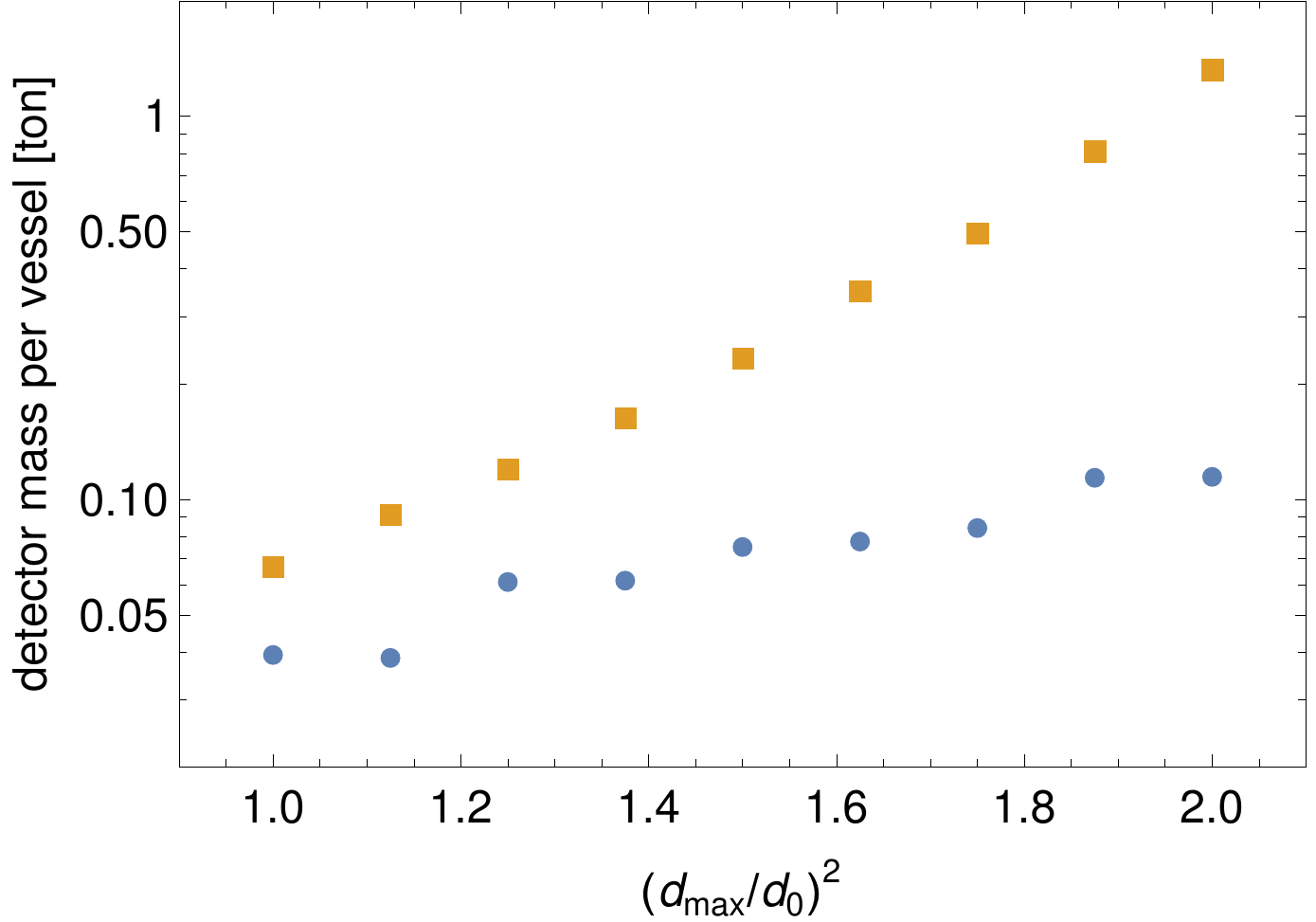}%
   
    \caption{Detector mass required to obtain a 90\% detection probability with a 5\% false positive rate as a function of the parameter $\xi=(d_\mathrm{max}/d_0)^2$, the average reactor power masking factor. We take $d_0=5\,$m. The blue circles are the result for a single vessel detector, whereas the orange squares show two vessels berthed side-by-side. }
    \label{fig:resx}
\end{figure}

Based on the event number for a fresh core (108\,ton$^{-1}$ at 5\,m distance) and background numbers (25\,ton$^{-1}$) we can
compute the required detector size to verify the presence of a
neutrino signal and, hence, the presence of a nuclear reactor which
produced appreciable power. We employ the usual likelihood ratio test, but will compute the probability distribution of the test statistics using Monte Carlo methods; for details see the appendices. We require a detection probability for a diversion (absence of a neutrino signal) of 90\% and, at the same time, require a false positive rate (reactor absence determined when a reactor is indeed present) of 5\%, as is safeguards practice~\cite{IAEA2001}. We also set as a figure of merit the required detector mass to attain this goal. The introduction of a variable distance as an information barrier makes the analysis more complicated and we profile the likelihood over the specific distance range corresponding to the desired power masking factor, $\xi$. The result for a single submarine is shown in Fig.~\ref{fig:resx} as blue circles. The case $\xi=1$ corresponds to no information barrier and no distance variation and we find that, over the whole range for $\xi$, detectors of around 100\,kg mass can achieve this measurement. We label this deployment option {\it single boat detector}.

However, since we are potentially dealing with several, mobile reactors at once (one per boat) the question of spoofing the observed submarine signal using additional reactor cores arises. As one of the most difficult cases, we consider two submarines berthed side-by-side for the full duration of the measurement. The geometry of this situation is depicted in the left hand panel of Fig.~\ref{fig:port}. We now have two reactors whose signal strengths are unknown and two distances that can each vary independently within $\xi$. The results are shown as orange squares in Fig.~\ref{fig:resx} and the details of the calculation and their game-theoretical explanation can be found in the appendix. For $\xi=1$, no information barrier, the increase in detector mass is modest, but this increase grows to more than a factor 10 for $\xi=2$. Here we assume that both vessels have a fresh core; the increase would be even larger if the vessel under inspection has a fresh core, but the side-by-side vessel has an older core. Thus, unconstrained side-by-side berthing ({\it i.e.}, no restrictions on the vessels that can be berthed together) will require many larger detectors, unless $\xi$ can be kept below 1.5. Alternatively, constrained side-by-side berthing ({\it e.g.}, where the duration of berthing or age of nearby reactor cores is prescribed) imposes operational constraints and requires unique, reliable hull identifiers with an associated verification system, but keeps the detector masses smaller while still requiring multiple detectors. 

At this point, it is worthwhile to look beyond the individual vessel to the overall fleet: If we indeed require $\xi=2$ it may become more effective and impose fewer operational constraints to deploy a single, larger detector at a dedicated measurement site that can perform the measurement in a short time on all submarines in the fleet, one at a time. Here, operationally, the vessel berths at the detector site by itself, but only for a few days. As we can infer from the single vessel results, a 1\,ton detector without interference from the backgrounds of a side-by-side vessel could achieve a measurement in 6 days. This measurement time shortens as more patrols are conducted, since cerium eventually reaches its equilibrium abundance. Already after the third patrol the time required will be of the order of a day. Thus, a single ton-scale detector could serve a full fleet of vessels, an option we designate as a {\it whole fleet detector}.

We find for all cases considered that for a single boat detector a 100\,kg
mass detector would achieve such a measurement after the
first patrol tour using a fresh core, even in the presence of
realistic backgrounds. Also, in considering a worst case spoofing scenario we find that it can be countered by increasing the detector mass for each vessel to 1--2\,ton for side-by-side berthing. Thus,
the main safeguards objective, verification of the vessel being
nuclear-powered, can be met by CeRuLEAN under a wide range of circumstances: CeRuLEAN can effectively
fulfill the same role as the flux tabs proposed in Ref.~\cite{shea},
but without the need to obtain access to the vessel. In addition, we examined the prospect of verifying core enrichment levels using CeRuLEAN and find it is also possible for much larger detector mass of around 10\,tons; for calculations and details see the appendices.

A notional concept of operation could look like this: the inspecting
party delivers a detector to the inspected party and oversees its
installation at the submarine's berthing site, while the submarine is
at sea. The inspected party verifies that the detector can indeed only
detect neutrinos. The detector is then installed in a manner which
allows the inspected party to vary the reactor-detector distance, from
say 5--7.1\,m ($\xi=2$), without the actual reactor-detector distance being
ascertainable by the inspecting party.

Installing a movable detector would be a one-time effort and would not
require re-installation each time the detector distance needed to be
changed. For example, an IBD-based, segmented plastic scintillator
detector, located on a movable platform whose position is altered on
average three times per week, has been in operation at a 3\,GW$_\mathrm{th}$ power
reactor as part of the DANSS neutrino experiment for some years
~\cite{DANSSpre, DANSSproc}.

When a submarine returns from patrol the vessel docks at its berthing
site as usual and the proper detector-reactor distance is chosen so as
to mask the FPED accrued on that patrol. Subsequently, the neutrino
detector records its signal and transmits the resulting data to the
inspecting party. In this naval reactor safeguards scheme using
CeRuLEAN, no alteration to submarine operations itself is required and
no access to the submarine or any patrol data is necessary.

\section{Summary}

We propose a new technique to determine the presence and fuel enrichment level of a
nuclear reactor, called the CErium RUthenium Low Energy AntiNeutrino
measurement, or CeRuLEAN. In this paper, we studied the application of
CeRuLEAN to the problem of safeguarding naval reactor fuel in military
applications, specifically, nuclear-powered submarines. The CeRuLEAN
method uses antineutrino emissions from long-lived fission products in
a shutdown reactor. It is an offline method and a measurement that can
be performed while a submarine is in port. The required IBD neutrino
detector would be similar to demonstrated
prototypes~\cite{Haghighat:2018mve} and would employ a segmented
plastic scintillator design. Scaling from existing measured
backgrounds appears to yield manageable background rates.

CeRuLEAN measurements can be used to verify two aspects of naval reactor declarations.  First, they can verify the presence of
a nuclear reactor (that has been producing more than a certain amount
of energy) for a fresh core after just the
first submarine patrol with a small detector of about 100\,kg mass in a single boat detector deployment. We also considered spoofing scenarios and find that they can be dealt with using a 1-2\,ton detector per boat for vessels berthed side by side. We also find that a 1-2\,ton detector can serve as a whole fleet detector if prolonged, close proximity (less than 20\,m) to other vessels is prevented. This shows that CeRuLEAN can directly replace neutron flux tabs~\cite{shea} under a wide range of deployment scenarios. Second, CeRuLEAN could verify the type of enriched core present, LEU or HEU, however requiring overall larger detectors of order 10\,tons. 

A key advantage of CeRuLEAN is that the neutrino signal can only be spoofed by other reactors, since it originates from the decay of
$10^{16}-10^{18}\,$Bq of $^{144}$Ce/$^{106}$Ru. Thus, it provides an
alternative to the naval flux monitors proposed in Ref.~\cite{shea}, which would likely have to be mounted in the reactor
compartment. Therefore, their placement and retrieval requires
on-board access to the most sensitive parts of the vessel. In
contrast, the naval reactor verification scheme based on the
CeRuLEAN method presented here incurs no significant
operational encumbrances and discloses no sensitive reactor design or
operation information, beyond the core plutonium fission fraction. Most importantly, in this applied physics context CeRULEAN requires no on-board access to the military submarine by civilian inspectors to verify reactor declarations. Therefore, CeRuLEAN may provide the first cross-over technology transfer opportunity for antineutrino detectors from high energy physics to timely non-proliferation technical challenges.

\section*{Acknowledgements}

We would like to thank Thomas E. Shea for valuable comments on earlier
versions of the manuscript.  The work was supported by the
U.S. Department of Energy Office of Science under award number
DE-SC00018327 and by the National Nuclear Security Administration
Office of Defense Nuclear Nonproliferation R\&D through the Consortium
for Monitoring, Technology and Verification under award number
DE-NA0003920.

\bibliography{apssamp}

\clearpage

\begin{appendix}

\section{Statistical analysis}  

The IAEA aims for a 90\% detection probability with a 5\% false positive rate~\cite{IAEA2001} and, thus, we will use these as benchmarks. The statistical problem can be cast as a hypothesis test: The null hypothesis H$_0$ is that there is no reactor present, whereas, the alternative hypothesis H$_1$ is that there is a reactor present. In practice, a neutrino observation will provide a measurement of the signal strength parameter $x$, where we define $x=0$ as the background-only case and $x=1$ as a reactor with nominal FPED. The continuous parameter $x$ allows H$_0$ to be transformed into H$_1$ and vice versa; note that $x\in\mathbb{R}_0^+$, that is $x$ can not be negative. We are dealing with a so-called nested hypothesis test. As a test statistic we will use the maximum likelihood estimate for $x$, called $\hat x$. For the likelihood we use the usual Poisson likelihood, since we are dealing with a counting experiment with a small number of counts:
\begin{equation}
-2 \log L= 2\,\sum_{i=1}^{N} \left(\bar n_i - n_i\right)+n_i\log \frac{n_i}{\bar n_i}\,,
\end{equation}
where $N$ is the number of energy bins, $n_i$ is the data in bin $i$ and $\bar n_i$ the prediction for $n_i$. The model for the prediction is given as:
\begin{equation}
    \bar n_i = x\,\frac{d^2}{d_0^2}\,n_i^0 + b_i\,,
\end{equation}
where $n_i^0$ is the number of expected events at distance $d_0$ for a patrol at nominal mean reactor power, $d$ is the unknown actual reactor-detector distance and $b_i$ is the expected number of background events in bin $i$. We assume that the background is constant with energy and well known from measurement periods without a vessel present. The distance $d$ is treated as a nuisance parameter and profiled over the appropriate range to achieve the desired power masking factor $\xi$. In all calculations $x\geq 0$.

The task is now to find a critical value $x_c$, where we accept H$_0$ if $\hat x \leq x_c$ and reject it otherwise.  We need to chose $x_c$ in such a way that we obtain a false positive rate, {\it i.e.} we accept H$_0$ when H$_1$ is true, of 5\% (or less) and a detection rate, {\it i.e.} we accept H$_0$ when H$_0$ is true, of 90\% (or more). Usually, one can use Wilks' theorem to compute the distribution function of a test statistic based on maximum likelihood estimates. However, in our case the fact the we are dealing with small count rates and a physical boundary at $x=0$ makes it preferable to resort to a direct Monte Carlo computation. We take either H$_0$ or H$_1$ to be true, {\it i.e.} $x_0=0$ or $x_0=1$, respectively, and compute the expected number of signal plus background events in each energy bin. We then impose Poisson-distributed fluctuations, determine the maximum likelihood estimate $\hat x$ and repeat this exercise 10,000 times to obtain a numerical realization of either $p(\hat x|\mathrm{H}_0)$ or $p(\hat x|\mathrm{H}_1)$. The conditions for $x_c$ then are
\begin{eqnarray}
    \int_0^{x_c} d\hat x\,\, p(\hat x|\mathrm{H}_0)&\leq&1-0.9\,,\nonumber\\
    \int_{x_c}^\infty d\hat x\,\, p(\hat x|\mathrm{H}_1)&\geq&1-0.05\,.
\end{eqnarray}
In practice we solve the second condition for the equal sign and check if the first condition is fulfilled, {\it i.e.} we fix the false positive rate to 5\% and consider configurations with a detection rate of 90\% or more adequate. 

\section{Game theory of power masking}

Power masking is achieved by allowing the host to vary the reactor-detector distance $d$ within a preset range. In the right hand panel of Fig.~\ref{fig:port} we show  one possible arrangement allowing the standoff distance between reactor core and detector to be changed without changing the depth of the detector and, hence, background. This is important to prevent the inspecting party from inferring the actually chosen distance $l_i$ (and hence the submarine's FPED during a given patrol), which is given as $l_i^2=y_i^2+x_i^2$. The actual choice of $d$ is a free parameter for the host and for the generation of Monte Carlo truth data. In the analysis we have to account for the ignorance of the inspecting party of this actual value by profiling over the distance. This hypothesis-testing scenario can, alternatively, be understood using the terms of game theory.  As stated, the inspected party is free to choose a a distance for each detector $d_i$ from its  reactor $R_i$ in the range $d_0$ to $d_\mathrm{max}$. We find that both the detection probability and false positive rate are monotonic functions of $d_0$ and $d_\mathrm{max}$. Since we have a requirement to keep the false positive rate below 5\%, the inspected party can exploit this by choosing an arrangement that forces us to set $x_c$ as high as possible, thereby reducing our detection probability. The inspectorate, on the other hand, has to assume the worst case for both hypotheses, H$_0$ and H$_1$, in the analysis. In game theory terms this problem is a finite, simultaneous\footnote{In reality, the moves of each party happen in sequence, but since neither party knows the move made by its opponent this reduces to a simultaneous game.}, non-cooperative game: We have two players, the host country (C) and the inspectorate (I) and the cost function is based on the detection probability for a diversion at fixed false positive probability. The cost of a false positive is much lower than the cost of a false negative, {\it i.e.} an undetected diversion, but is difficult to quantify in either case. Instead, we stipulate the following: in this game player I wants to ensure that a given detection probability can be obtained irrespective of the strategy player C has chosen and under the most conservative analysis assumptions.

Specifically, player C can choose either H$_0$ (diversion) or H$_1$ (no diversion). If H$_1$ (no diversion) is chosen then the game is moot. Thus, we assume player C chooses H$_0$ (diversion) and can next choose distances for the detector to be either n (near) or f (far), where near corresponds to $d_0$ and far to $d_\mathrm{max}$. Thus, there are 2 available strategies for player C. Player I needs to find $x_c$ for which the false positive rate is 5\%, which is done by assuming  H$_1$, and here player I has the same choices for the distance $d$ as above. To then evaluate $p_D$, player I has to additionally assume H$_0$ and, in doing so, has again two choices for the distance, resulting in altogether 4 strategies available to player I. Thus, the most general payoff matrix has $2\times4$ entries.
Instead of a payoff matrix, the information is easier to show in the following form, where capital letters denote the strategy of player C, lower case letters the strategy of player I and the subscripts 0 and 1 indicate player I's choices for the evaluation of $p(\hat x|H_{0/1})$, respectively. The following is one example of the resulting detection probabilities, for the single boat detector case, for a specific set of distances and detector masses:
\begin{equation*}
\begin{array}{l|l}
\begin{array}{ccc}
 \text{N} & \text{n}_1 & \text{f}_1 \\
 \text{n}_0 & 1. & 1.  \\
 \text{f}_0 & 1. & 1. \\
\end{array}&%
\begin{array}{ccc}
 \text{F} & \text{n}_1 & \text{f}_1 \\
 \text{n}_0 & 0.98 & 1 \\
 \text{f}_0 & 0.93 & 0.98  \\
\end{array}
\end{array}
\end{equation*}
The worst case scenario for player I occurs for the combination F f$_0$ n$_1$, which is consistently true for all cases studied. This result can be understood as follows: if we neglect the effect of limited count rate statistics, the $\hat x$ of our maximum likelihood estimate will be approximately $x$ where $x=0$ for H$_0$ and $x=1$ for H$_1$. By choosing detector configuration F player C moves the distribution of $\hat x$ to smaller values than by choosing N: for a given real signal $x$ we obtain fewer counts by choosing F. By choosing f$_0$ player I moves the distribution of $\hat x$ to larger values under H$_0$, that is the same number of counts requires a larger signal strength $x$ for the choice f. Conversely, player I in choosing n$_1$ moves the distribution of $\hat x$ to smaller values under H$_1$, that is the same number of counts requires a smaller signal strength $x$ for the choice n. Thus, this combination of choices requires the largest possible value of $x_c$ and, hence, results in the smallest possible detection probability at fixed false positive rate.

In port, submarines can be berthed either individually or side by side. In the main text we have analysed the case of an individually berthed submarine in detail. For the side-by-side berthing (SBSB) case shown in the left panel of Fig.~\ref{fig:port}, several other geometric factors come into play, notably the gap $g$ between two vessels. Also, we now have to consider the signal of the reactors $R_{i-1}$ and $R_{i+1}$ of the neighbouring vessels, with distances $l_{ij}$ between detector $i$ and reactor $j$ like this
\begin{equation}
    l_{ij}^2=(i-j)^2(b+g)^2+y_i^2+x_i^2\,,
\end{equation}
which for $i=j$ reproduces the result for $l_i$ as given above.

The presence of a second vessel, and thus reactor, relatively close to the vessel under inspection (VUI) greatly complicates the analysis, in particular since it allows for active spoofing of the neutrino signal from the VUI. Moreover, the spoofing potential is increased considerably by the variable distance information barrier or power masking factor.  It is easy to show that for a single detector in this situation a complete degeneracy between the signal from the second vessel and the VUI exists and, thus, no conclusion about the VUI can be drawn. Therefore, we need one detector for each vessel in an arrangement where each detector is considerably closer to one of the two vessels, as shown in the left hand panel of Fig~\ref{fig:port}.

Again, the inspected party is free to choose a a distance for each detector in the range $d_0$ to $d_\mathrm{max}$ and player C can choose either diversion H$_0$ or no diversion H$_1$; but, since no diversion leads to a moot game, we assume player C chooses H$_0$ (diversion). Now player C can choose distances for detector 1 and 2 from these 4 combinations (nn), (nf), (fn), (ff) where n stands for near, $d_0$, and f stands for far, $d_\mathrm{max}$, for a total of 4 available strategies. Player I needs to find $x_c$ for which the false positive rate is 5\%, which is done by assuming  H$_1$, and player I has the same four choices for distance pairs as above. To then evaluate $p_D$, player I has to also assume H$_0$ and in doing so has again four choices for distance pairs, resulting in altogether 16 strategies available to player I. Thus, the most general payoff matrix has $4\times16$ entries.
Instead of a payoff matrix, as for the single boat detector example, the information is easier to show in the following form, where capital letters denote the strategy of player C and lower case letters and subscripts the strategy of player I. The following is one example for specific distances and detector masses in the side-by-side berthing case:
\begin{equation*}
\begin{array}{l|l}
\begin{array}{ccccc}
 \text{NN} & \text{nn}_1 & \text{nf}_1 & \text{fn}_1 & \text{ff}_1 \\
 \text{nn}_0 & 1. & 1. & 1. & 1. \\
 \text{nf}_0 & 1. & 1. & 1. & 1. \\
 \text{fn}_0 & 0.99 & 0.93 & 1. & 1. \\
 \text{ff}_0 & 1. & 0.99 & 1. & 1. \\
\end{array}&%
\begin{array}{ccccc}
 \text{NF} & \text{nn}_1 & \text{nf}_1 & \text{fn}_1 & \text{ff}_1 \\
 \text{nn}_0 & 1. & 1. & 1. & 1. \\
 \text{nf}_0 & 1. & 1. & 1. & 1. \\
 \text{fn}_0 & 0.99 & 0.98 & 1. & 1. \\
 \text{ff}_0 & 1. & 1. & 1. & 1. \\
\end{array}\\
\hline
\begin{array}{ccccc}
 \text{FN} & \text{nn}_1 & \text{nf}_1 & \text{fn}_1 & \text{ff}_1 \\
 \text{nn}_0 & 1. & 0.09 & 1. & 0.74 \\
 \text{nf}_0 & 1. & 0.36 & 1. & 0.99 \\
 \text{fn}_0 & 0.91 & 0.04 & 1. & 0.44 \\
 \text{ff}_0 & 1. & 0.535 & 1. & 0.999 \\
\end{array}&%
\begin{array}{ccccc}
 \text{FF} & \text{nn}_1 & \text{nf}_1 & \text{fn}_1 & \text{ff}_1 \\
 \text{nn}_0 & 1. & 0.87 & 1. & 1. \\
 \text{nf}_0 & 1. & 0.99 & 1. & 1. \\
 \text{fn}_0 & 0.83 & 0.31 & 1. & 0.99 \\
 \text{ff}_0 & 0.98 & 0.84 & 1. & 1. \\
\end{array}
\end{array}
\end{equation*}

We find that FN fn$_0$ nf$_1$ yields the worst detection probability, a mere 0.04, in this example. In all cases studied it is this combination, FN fn$_0$ nf$_1$, that results in the smallest detection probability. We also note that the detection probabilities for the cases FN fn$_0$ fn$_1$ and NN nn$_0$ nn$_1$ are uniformly high. This indicates the high price we pay, in the form of a reduced detection probability, for the use of distance as an information barrier. This result can be understood along the lines of the argument made for the single boat detector.
The choice FN by player C makes the distribution for $\hat x_1$ move to large values of $x_1$ and small values for $x_2$. The choice fn$_0$ by player I assumes detector 1 to be far and thus a large $x_1$ is compatible with a low count rate, whereas, assuming detector 2 to be near makes small $x_2$ more likely, thus driving the estimate for $x_1$ even higher. On the other hand, under the assumption of H$_1$ the choice nf$_1$ of player I renders the estimate for $x_1$ as small as possible, thereby maximising the overlap of the distributions of $\hat x$ under H$_0$ and H$_1$, respectively. The difference between the single boat and multiple boat detector cases is that the level of confusion is much greater and, thus, the detection probability plummets.

\section{Reactor details}

The parameters used in Ref.~\cite{Ippolito}, on which our models are based, are reproduced in Tab.~\ref{tab:cores}

\begin{table}[h]
    \centering
    \includegraphics[width=\columnwidth]{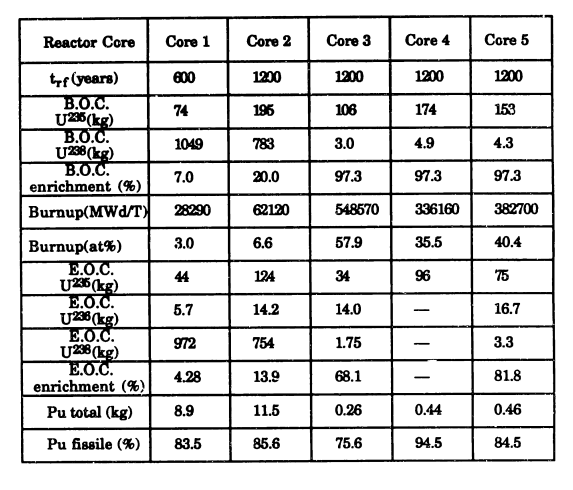}
    \caption{Table 5.5 reproduced from Ref.~\cite{Ippolito}. All variations studied were for a 50\,MW$_\mathrm{th}$ core
output. Core 1 is designated as LEU7, Core 2 as LEU20 and Core 4 as HEU.}
    \label{tab:cores}
\end{table}

The U.S. Navy has repeatedly~\cite{navy2014,navy1995} stated its preference for HEU cores because, inter alia, HEU allows for lifetime cores with no refueling. In this context, among the 97.3\% enrichment HEU model options in Tab.~\ref{tab:cores}, Cores 4 and 5 are the most suitable proxies, since they achieve their lifetime with a much lower burn-up than Core 3 and so pose fewer challenges in terms of fuel longevity. Thus, we choose Core 4 as our HEU core model. Core 2 represents an LEU option with the same nominal reactor lifetime as any of the HEU cores and, hence, could be a good candidate for a low-enriched replacement, which we designate as LEU20. Note, for a would-be proliferator this design is not attractive as it requires, overall, a higher uranium loading than any of the HEU cores\footnote{Enriching to 20\% requires nearly as much effort, in terms of separative work units, as enriching to 97\%.}. Core 1, which we will call LEU7, is similar in enrichment to what is reported for French and Chinese naval reactors~\cite{nti}, but has a much shorter lifetime. Given its sharply reduced uranium content, this is, in principle, an attractive option for a skilled proliferator, allowing diversion of up to 68\% of the original HEU core.

\begin{figure}
    \centering
    \includegraphics[width=\columnwidth]{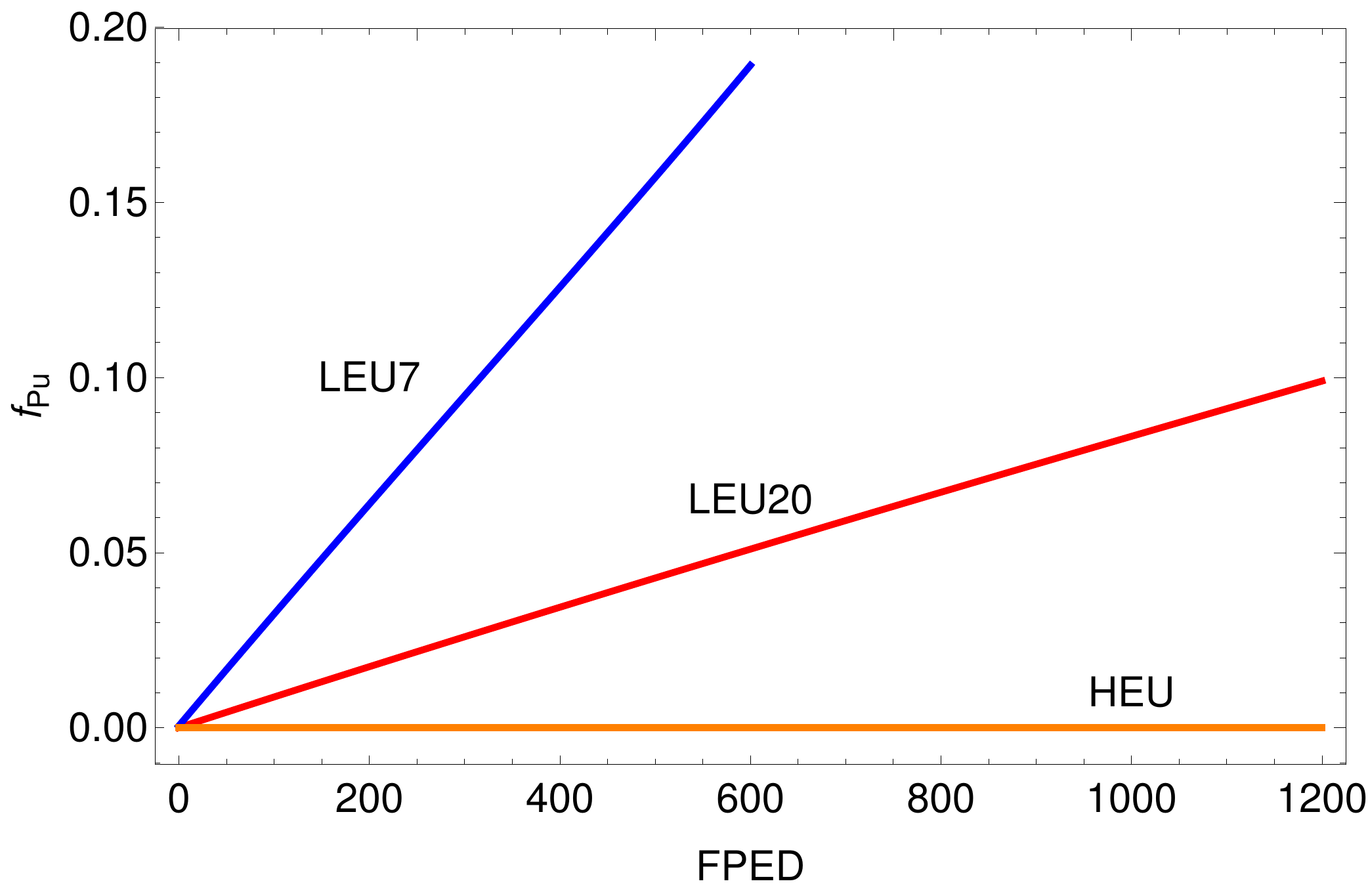}    \caption{Fission fractions of $^{239}$Pu for cores LEU7, LEU20 and HEU, as a function of full power equivalent days (FPED). Reconstructed from Ref.~\cite{Ippolito}.}
    \label{fig:ff}
\end{figure}

Finally, to obtain the required nominal reactor power of 150\,MW$_\mathrm{th}$, we have to scale the numbers in Tab.~\ref{tab:cores} by a factor of 3. Using Figs.~5.8, 5.9 and 5.11 of Ref.~\cite{Ippolito} we can reconstruct the fission fractions of $^{239}$Pu for LEU7, LEU20 and HEU which are shown in Fig.~\ref{fig:ff}

\section{Enrichment analysis}

As noted in the main text, measuring the ratio of {\ce} to {\ru} is a direct
measurement of the plutonium fission fraction. Furthermore, the
cumulative fission yield for $^{106}$Ru is 0.025 for fast neutron
fission of $^{238}$U~\cite{jendl}. However, $^{238}$U fission is also
highly correlated with plutonium production, since both reaction
pathways start with a neutron interacting with $^{238}$U. Therefore,
this would enhance the signature by as much as 50\%, depending on the
fission fraction of $^{238}$U. Unfortunately, Ref.~\cite{Ippolito}
does not provide enough information to infer this fission fraction, so
we do not consider it further here.

The following verification cases can be distinguished:
    \begin{enumerate}
        \item For an HEU core, we would like to verify that it is
          indeed an HEU core and it has not been replaced with an LEU
          core, in order to divert HEU. This diversion pathway does
          require significant nuclear expertise.
        \item For an LEU core, we would like to verify that it is
          indeed an LEU core. This case would arise should
          there be a treaty committing navies to use LEU cores, as for
          instance proposed in Ref.~\cite{shea}.
        \item For a combined agreement, such as beginning a program
          with existing HEU cores and eventually transitioning to LEU
          cores, we would like to verify that the change in core
          enrichment occurs as prescribed by the details of
          the specific agreement.
    \end{enumerate}

\begin{figure}
    \centering
    \includegraphics[width=\columnwidth]{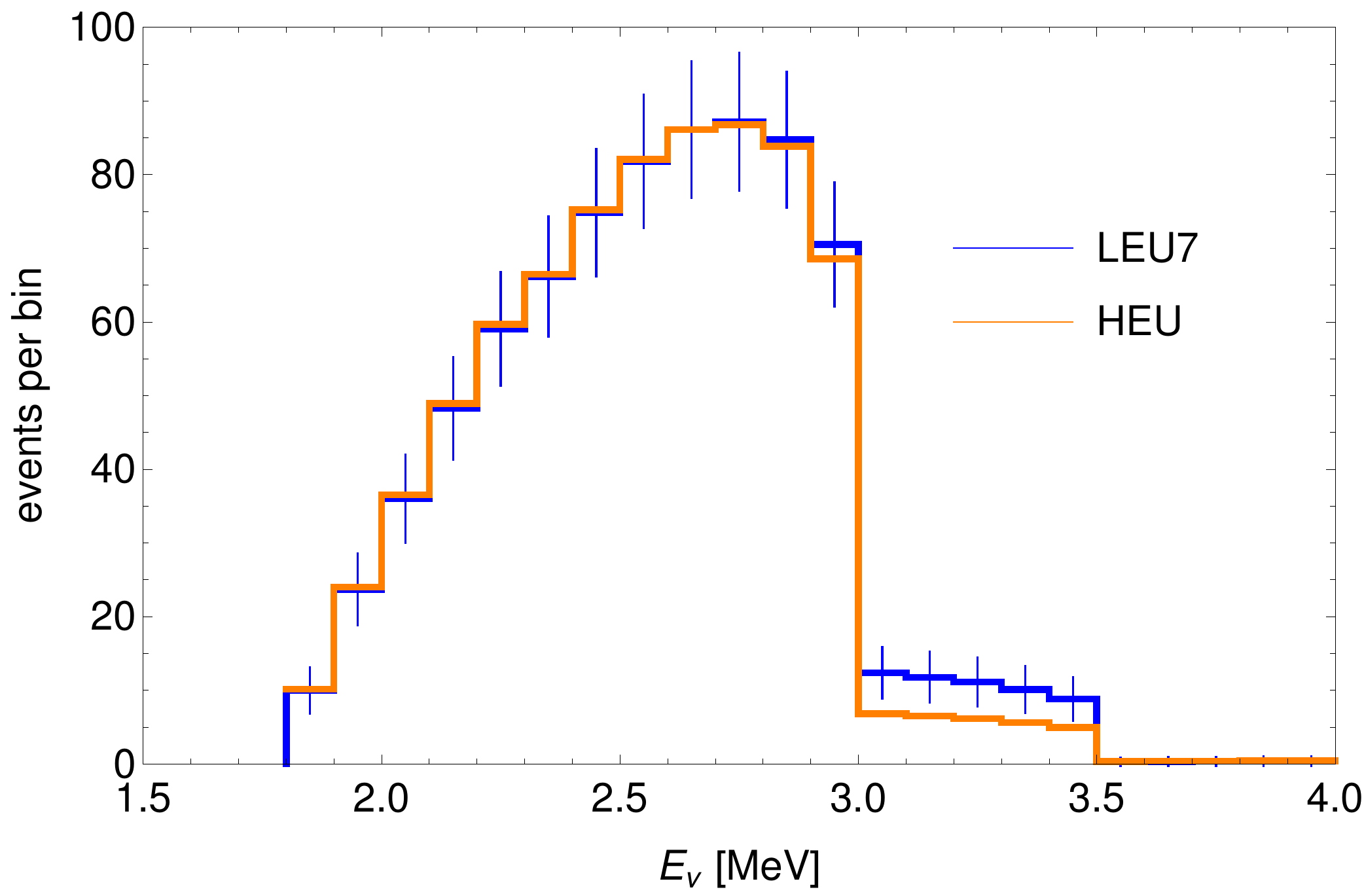}
    \caption{Shown are the event rates for a 2 month measurement in year 5 of reactor lifetime for cores HEU and LEU7, where the area of the curves below 3\,MeV has been adjusted to match.}
    \label{fig:ceru}
\end{figure}

For a determination of core enrichment, we need
to look at the energy spectrum of neutrinos. The resulting event rate
spectrum for a 10\,ton detector is shown in Fig.~\ref{fig:ceru}. The
difference between the orange and blue lines in the range 3 to
3.5\,MeV is what distinguishes an HEU core from an LEU core. More
importantly, this result is obtained while leaving the normalization
practically free with a 100\% uncertainty. We checked that changing the binning from 0.1\,MeV to 0.2\,MeV
only mildly affects the result. This range of bin widths corresponds
to an energy resolution of around $(6-12)\%/\sqrt{E}$, which has been
achieved with segmented neutrino
detectors~\cite{Ashenfelter:2018iov,Haghighat:2018mve}.

We did
study in detail the use of coherent elastic neutrino nucleus
scattering (CEvNS) and find that a 1\,ton tungsten detector with
10\,eV nuclear recoil threshold would yield results slightly less
sensitive than those presented here. However, the background question
would need significant study and R\&D. Thus, we do not further
consider this option for now.

\begin{figure}
    \centering
    \includegraphics[width=\columnwidth]{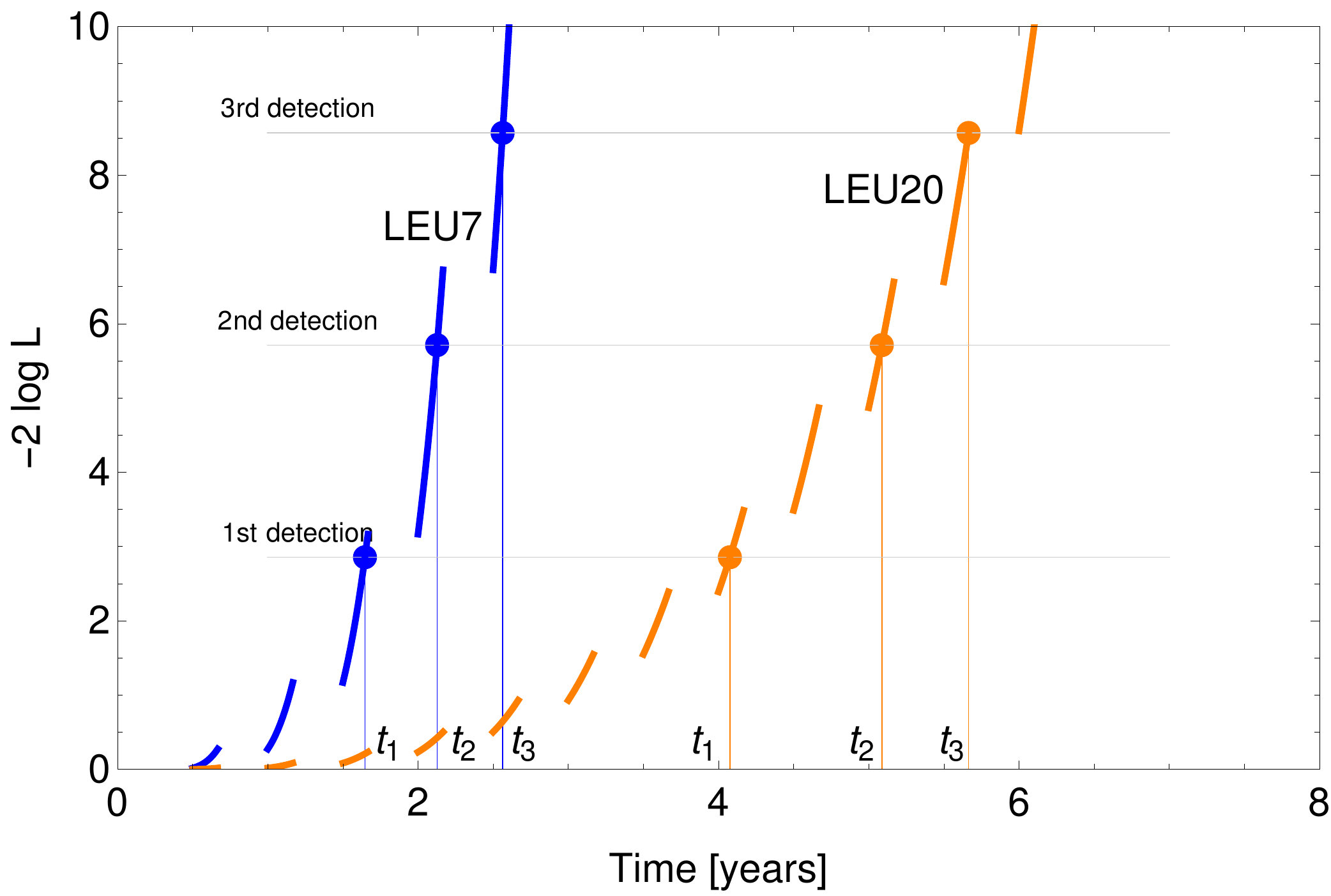}
    \caption{$-2 \log L$ as a function of calendar time for the two cases of distinguishing the LEU7 from an HEU core and the LEU20 from an HEU core. The gaps in the curve denote the times the vessel is at sea. The marked time $t_i$ corresponds to the $i$-th detection at 95\% confidence level.}
    \label{fig:chi}
\end{figure}

We can now proceed to compute the statistical significance with which we can distinguish the LEU7 and LEU20 cores from an HEU core. We set up the usual likelihood function and assume data taking occurs during each period in port. For each data taking period we allow the signal normalization to float freely with a 100\% uncertainty. We then add the likelihood $L$ from each data taking period. The result is shown in Fig.~\ref{fig:chi} as a function of calendar time.
The horizontal lines correspond to the $-2 \log L$ increase needed to obtain a 95\% confidence upper limit on the {\ru} contribution to the data and, hence, plutonium fission fraction $f_\mathrm{Pu239}$.

The time to first detection $t_1$ of a non-HEU core is 1.6 years for
the LEU7 core and 4.1 years for the LEU20 core. Subsequent
verification of the same reactor will be obtained much faster, as is
evident in Fig.~\ref{fig:chi}. This is due to the increased amount of
plutonium accumulated in a core over time due to fission (relative to
the beginning of life for a reactor when there is no plutonium
present), which occurs at naturally differential rates in cores of
different initial enrichment, {\it i.e.}, LEU7 versus LEU20. Time to
first detection for both cores corresponds in calendar years to about
1/5th of the service life of the reactor core. Therefore, we find that
CeRuLEAN can also meet the three safeguards cases identified at the beginning of this section. That
is, CeRuLEAN can be used to prove that a core is indeed HEU (case
1) or conversely to prove that it is not (case 2). Alternatively, the time to first detection of a non-HEU core by
CeRuLEAN could be used to confirm that a core has been converted from
HEU to LEU (case 3).  This capability exceeds what conventional
safeguards can achieve, especially in the case of already deployed
submarines where the fueling of the reactor was not subject to initial
safeguards.

To put the novel CeRuLEAN measurement in context, existing safeguards
practice at civilian power reactors is to apply seals and to inspect
the core at refueling to verify declarations and/or detect a
diversion. For a would-be proliferator, in the case of naval reactors,
only the LEU7 option would be relevant.  The CeRuLEAN 1.6 years (or 19
months) detection time is not dissimilar to the fuel cycle duration of
many civilian reactors. Thus, CeRuLEAN could provide an indication of
a diversion from a naval reactor on a time scale that is similar to
what is currently achieved for safe-guarded civilian nuclear reactors.

\end{appendix}

\end{document}